\documentclass[aps,prl,twocolumn,showpacs,preprintnumbers,amsmath,amssymb]{revtex4}
\usepackage{graphicx}
\usepackage{bm}
\usepackage{dcolumn}

\def\nat#1#2#3{Nature {\bf #1}, #2 (#3)}
\def\prl#1#2#3{Phys. Rev. Lett. {\bf #1}, #2 (#3)}

\def\pra#1#2#3{Phys. Rev. A {\bf #1}, #2 (#3)}

\def\epl#1#2#3{Europhys. Lett. {\bf #1}, #2 (#3)}

\def\pre#1#2#3{Phys. Rev. E {\bf #1}, #2 (#3)}
\def\prb#1#2#3{Phys. Rev.  B {\bf #1}, #2 (#3)}

\def\noi{\noindent}
\def\bc{\begin{center}}
\def\ec{\end{center}}
 \newcommand{\bea}{\begin{equation}}
 \newcommand{\eea}{\end{equation}\noi}
 \newcommand{\ber}{\begin{eqnarray}}
 \newcommand{\eer}{\end{eqnarray}\noi}
\begin{document}
\title{Proposed fitting function for the critical Casimir force on $^4$He film below the $\lambda$ point}
\author{Shyamal Biswas}\email{sbiswas.phys.cu@gmail.com}

\vskip0.4cm \affiliation{Department of Theoretical Physics, Indian Association for the Cultivation of Science, Jadavpur, Kolkata 700032, India}

\date{\today}

\begin{abstract}
We have extended the mean field calculation \cite{kardar} of Zandi et al and have obtained an approximate mathematical expression for the Casimir scaling function $\vartheta(y)$, which if extrapolated to the domain $\pi^2\ge-y\ge 0$, becomes surprisingly similar to that obtained experimentally \cite{chan} by Ganshin et al. The extrapolated $\vartheta(y)$ can be regarded as a proposed fitting function, that appears to agree better with the experiments \cite{chan} in D=3, than the exact mean field result of Zandi et al \cite{kardar}.
\end{abstract}
\pacs{67.25.dj, 67.25.D-, 67.25.dp, 05.70.Jk}

\maketitle
\subsection{1. Introduction}
It was proposed two decades ago by Nightingale and Indekeu that, the confinement of the critical fluctuations may give rise to a (classical) Casimir force \cite{nightingale}. Thereafter an experimental verification of the same by Mukhopadhyay and Law \cite{mukhopadhyay} made the study of the classical Casimir force on critical films a hunting ground to the experimentalists and theoreticians. On this issue, the Casimir effects on different critical films have been the subject of a number of experimental \cite{mukhopadhyay,chan2,chan-he3,fukuto,rafai,hertlein} and theoretical \cite{dietrich-he3,kardar,hucht,vasilyev,maciolek,diehl,hasenbusch,dohm} works within the last few years.

Few years ago Garcia and Chan \cite{chan2} and Ganshin et al \cite{chan} investigated the temperature ($T$) dependence of the critical Casimir force, and measured the Casimir force induced thinning of the liquid $^4$He film near the bulk $\lambda$ point ($T_\lambda=2.1768~\text{K}$). They obtained a universal scaling function ($\vartheta$) of the Casimir force, and observed a dip minimum in the Casimir scaling function below the $\lambda$ point. This experiment challenges our understanding of the finite size effects on the films near their bulk critical points.

For $T>T_\lambda$, a renormalization group calculation for the $\vartheta$ of the $^4$He film was nicely presented by Krech and Dietrich \cite{krech}. Although the calculation of Krech and Dietrich \cite{krech} matches well with the experiment \cite{chan2}, yet no satisfactory theory was given so far for $T<T_\lambda$. However, the mean field calculation of Zandi et al \cite{kardar} for this region, represents the basic feature of the classical Casimir force.

For $T<T_\lambda$, Zandi et al \cite{kardar} obtained an analytic expression of $\vartheta$ in terms of the maximum ($\phi_0$) of the superfluid order parameter ($\phi$). It is necessary to know $\phi_0$ for plotting the mean field part of the Casimir force. Although the graphical solutions of the $\phi_0$ are exact, yet the graphical solutions do not appear in the closed form. With a motivation of getting an approximate expression of $\phi_0$ in the closed form, we extend the calculation of Zandi et al \cite{kardar}.

Since the maximum correlation length ($\xi$) of a critical film (of thickness $L$) is $L/\pi$ \cite{kardar}, the order parameter appears into the film for $\xi<\frac{L}{\pi}$, and it vanishes for $\xi\ge\frac{L}{\pi}$ \cite{kardar}. Although our approximate mathematical expression of $\vartheta$ for $\xi<\frac{L}{\pi}$, matches very well with the exact mean field result of Zandi et al \cite{kardar}, yet we extrapolate our approximate expression of $\vartheta$ to the region $\xi\ge\frac{L}{\pi}$ where the order parameter actually vanishes. This extrapolation surprisingly fits the experimental data obtained by Ganshin et al \cite{chan}, and appears to agree better with the experiments \cite{chan}, than the exact mean field result of Zandi et al \cite{kardar}.

\subsection{2. Exact mean field result}

Before going into the details of our work, let us briefly reproduce the calculation of Zandi et al \cite{kardar}. For $T<T_\lambda$, we start from the Ginzburg-Landau free energy
 \bea
 F=A\int_0^L\bigg\{\frac{1}{2}\bigg(\frac{\phi(z)}{dz}\bigg)^2+\frac{r}{2}\phi^2(z)+u\phi^4(z)\bigg\}dz,
 \eea 
where $r=-1/\xi^2$, $\xi=\xi_o|T/T_\lambda-1|^{-\nu}$, $\nu$ is the correlation length exponent, $L$ ($\sim300\text{\AA}$ \cite{chan}) is thickness of the film along the $z$ direction, $A$ ($\sim1~\text{inch}^2$ \cite{chan}) is the area of the film along the $x-y$ plane, $u$ is a positive coupling constant, and the mean field order parameter ($\phi(z)$) obeys the equation \cite{kardar}
  \bea
-\frac{d^2\phi(z)}{dz^2}+r\phi(z)+4u\phi^3(z)=0.
  \eea
Solutions of the Eqn.(2) with the Dirichlet boundary conditions ($\phi(0)=\phi(L)=0$) are the Jacobian elliptic functions \cite{carr}. For the ground sate, there is a single maximum of $\phi(z)$ at $z=L/2$. With the consideration of Dirichlet boundary conditions ($\phi(0)=\phi(L)=0$), and that of a single maximum of $\phi(z)$ at $z=L/2$, we can write a transcendental equation for the maximum of the order parameter ($\phi(\frac{L}{2})=\phi_0$) as \cite{kardar}
\begin{eqnarray}
\sqrt{1-\eta}\sqrt{-y}/2=K\big(\eta/(1-\eta)\big),
\end{eqnarray}
where $y=L^2r$, $\eta=\frac{2u\phi_0^2}{-r}$, and $K(m)=\frac{\pi}{2}[1+\frac{m}{4}+\frac{9}{64}m^2+\frac{25}{256}m^3+...]$ is the complete elliptic integral of first kind. Since, the integration limit and $\phi(z)$, both are functions of $L$, the derivative of $F$ (in the Eqn.(1)) with respect to $L$ is nontrivial. The mean field force acting on the film, can be shown to be $-\frac{\delta F}{\delta L}=A\frac{1}{2}\big(\frac{d\phi}{dz}\big)^2\big|_{z=L}$ \cite{kardar}. Subtracting the bulk force ($-\frac{\delta F}{\delta L}\big|_{L\rightarrow\infty}$) from $-\frac{\delta F}{\delta L}$, we can obtain the mean field Casimir force, which can be scaled by $L^4/(\xi_oAk_BT_\lambda)$ to obtain the Casimir scaling function in the dimensionless form as \cite{kardar}
 \bea
\vartheta(y)=-\frac{y^2}{4\xi_ouk_BT_\lambda}\bigg(\frac{1}{4}-\eta(\sqrt{-y}/2)\big(1-\eta(\sqrt{-y}/2)\big)\bigg).
 \eea

\begin{figure}
\includegraphics{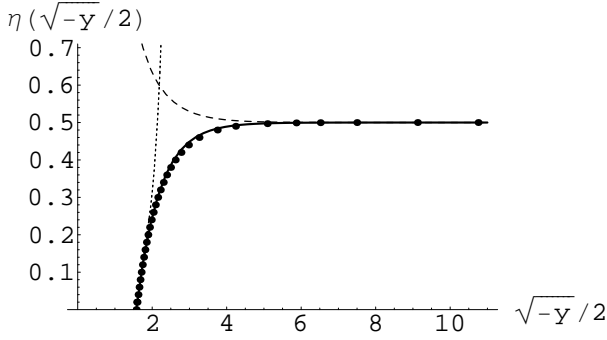}
\caption{The dots are the graphical solutions of the Eqn.(3). The dotted, dashed and continuous lines follow the Eqns.(5), (6) and (7) respectively.}
\end{figure}

To plot the scaling function (in the Eqn.(4)) we must need to know $\eta(\sqrt{-y}/2)$ from the Eqn.(3). Since $\eta$ by the definition is positive, one can check from the Eqn.(3) that, a non zero positive $\eta(\sqrt{-y}/2)$ exists only for \cite{kardar} $-y>\pi^2$. Although Zandi et al \cite{kardar} plotted the Eqn.(4) for $\infty>-y\ge 0$, yet they did not clearly describe how they obtained $\eta(\sqrt{-y}/2)$ for $\infty>-y>\pi^2$. However, one can easily check that $\eta(\sqrt{-y}/2)$ be exactly solved from the Eqn.(3) by the standard graphical method. Within this method, we can plot the left and right hand sides of the Eqn.(3) with respect to $\eta$ keeping a given $\sqrt{-y}/2$ fixed. The intersecting point of the two plots is a solution ($\eta$) with respect to the given $\sqrt{-y}/2$. In this way, we can get the set $\{\eta\}$ for a given set $\{\frac{\sqrt{-y}}{2}\}$. We plot these sets of points in the Figure 1. Adjusting the depth ($\vartheta(-7.39)=-1.3\pm0.03$ \cite{chan}) of the experimental minimum at $y=-\pi^2$ with a suitable value of $u$, one can plot the Eqn.(4) (with the graphical solutions) to the domain $\infty>-y>\pi^2$ for obtaining the $\vartheta(y)$ almost similar to that of Zandi et al \cite{kardar}. For $\pi^2\ge-y\ge 0$, Zandi et al \cite{kardar} plotted the Eqn.(4) by putting $\eta(\sqrt{-y}/2)=0$.

\subsection{3. An approximate mean field result in a closed form}

Although the graphical solutions do not give a mathematical expression of $\eta$ in terms of $\sqrt{-y}/2$, yet we need a mathematical expression of $\eta$ for conveniently plotting the Eqn.(3). This purpose can approximately be solved by interpolating the asymptotic solutions near $\eta=0$ and $\eta\rightarrow\frac{1}{2}$. It is easy to check from the Eqn.(3) that, the allowed range of $\eta$ is $0\le\eta<\frac{1}{2}$. For $\eta\rightarrow 0$, $\sqrt{-y}/2$ goes to $\pi/2$, and we can expand the $\eta$ parts of the Eqn.(3) up to the third order in $\eta$ to get
\begin{eqnarray}
&&\eta(\sqrt{-y}/2)=\frac{2}{3\pi^2}(-y-\pi^2)\big(1-\frac{25}{24\pi^2}(-y-\pi^2)\nonumber\\&&+\frac{1.04514}{\pi^4}(-y-\pi^2)^2+...\big)\ \ \text{for}  \ \  \frac{-y-\pi^2}{\pi^2}\ll 1.
\end{eqnarray} 
In the other asymptotic limit, i.e. for $\eta\rightarrow\frac{1}{2}$, $\sqrt{-y}/2$ goes to infinity. In this limit, $\eta$ can be expressed as $\eta=\frac{1}{2}(1-\delta)$, where $\delta\rightarrow 0$. In the Eqn.(3), $K(\eta/(1-\eta))$ can be written (by the definition) as an integral of the form $\int_0^1\frac{dp}{\sqrt{p(2-p)(1-\frac{1-\delta}{1+\delta}(1-2p+p^2)})}$, which tells that $p=0$ has a logarithmic divergence, so that most of the integral would come from $p\rightarrow 0$. In this asymptotic limit, the Eqn.(3) can be approximated with the first order in $\delta$ and $p$, as $\frac{\sqrt{-y}}{2}=\int_0^1\frac{dp}{\sqrt{2p(\delta+p)}}=\sqrt{2}\text{sinh}^{-1}[\frac{1}{\sqrt{\delta}}]$, which asymptotically gives
\begin{eqnarray}
\eta(\sqrt{-y}/2)=\frac{1}{2}\text{coth}^2\bigg(\frac{\sqrt{-y/2}}{2}\bigg) \ \ \text{for}  \ \  \frac{-y-\pi^2}{\pi^2}\gg 1.
\end{eqnarray} 

Let us now find a smooth function for $\eta$ as an interpolation for the whole range $0\le\eta<\frac{1}{2}$, in such a way, that, it fits to the two extreme ends of $\eta$. Since $\phi(z)=\sqrt{\frac{-r}{4u}}\text{tanh}[z\sqrt{-r/2}]$ is a solution of the Eqn.(2) for an unbounded situation, we can take a trial function from the asymptotic solutions (Eqn.(5) and (6)), that $\phi_0$ for the bounded system considered by us, would be close to $\sqrt{\frac{-r}{4u}}\text{tanh}\big(\frac{\sqrt{(-y-\pi^2)/2}}{2}\big)$. With this consideration, we can take the trial interpolation
\begin{eqnarray}
\eta(\sqrt{-y}/2)=\frac{1}{2}\text{tanh}^2\bigg(\frac{\sqrt{(-y-\pi^2)/2}}{2}\bigg)
\end{eqnarray} 
for the whole range $0\le\frac{-y-\pi^2}{\pi^2}<\infty$. In the Figure 1, we see that, the trial interpolation in the Eqn.(7), fits very well with the asymptotic solutions ($\eta(\sqrt{-y}/2)$) in the Eqn.(5) and (6). The trial interpolation also fits very well with the exact graphical solutions of the Eqn.(3). Hence, the approximate Eqn.(7) can be used for plotting the Eqn.(4).

\subsection{4. Extrapolation of the mean field result}

\begin{figure}
\includegraphics{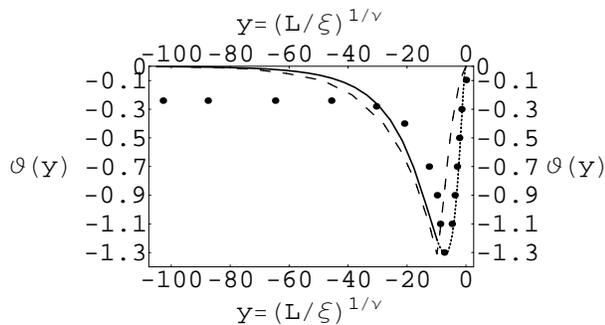}
\caption{The continuous part ($-y>\pi^2$) of the plot follows Eqn.(4), and the dotted part ($\pi^2\ge-y\ge 0$) is the extrapolation of the same equation. The depth of the experimental minimum \cite{chan} has been fitted for $\frac{1}{16u\xi_ok_BT_\lambda}=0.01237$. A few number of experimental points are replotted from the Ref.\cite{chan} with the substitutions $y=x/\xi_o^{1/\nu}$, $\nu=0.67016$ \cite{chan} and $\xi_o=1.2\text{\AA}$ \cite{kardar,ihas}. The dashed line is redrawn like that obtained by Zandi et al \cite{kardar}.}
\end{figure}

Since the minimum of the experimental $\vartheta$ is obtained at $y=-7.39\pm0.61$ \cite{chan,kardar}, and not at $y=-\pi^2$ \cite{kardar}, we extrapolate the Eqn.(4) (by sacrificing the fact that $\eta(\sqrt{-y}/2)=0$) to the domain $\pi^2\ge-y\ge 0$, with the motivations, whether the Eqn.(4) has a natural minimum near $y=-7.39$, and whether the experimental points be fitted with the extrapolation.

Putting the Eqn.(7) in to the Eqn.(4), we extrapolate the Eqn.(4) to the domain $\pi^2\ge-y\ge 0$ for obtaining the Casimir scaling function (in the Figure 2) similar to that obtained experimentally in the Ref.\cite{chan}. Now, we see in the Figure 2 that for $\xi_o=1.2\text{\AA}$ \cite{kardar,ihas}, Eqn.(4) has a natural minimum at $y=-7.03$, which is closer to the position ($y=-7.39\pm0.61$ \cite{chan,kardar}) of the experimental minimum. The extrapolated part surprisingly fits the experimental data of Ganshin et al \cite{chan}, and it appears to fit better than the exact mean field result of Zandi et al \cite{kardar}.

\subsection{5. Conclusions}

The extrapolation of the mean field result to the domain $\pi^2\ge-y\ge 0$ is questionable due to the fact that, $\eta$ is not being equal to zero, rather being negative in this domain. For this reason, the extrapolation does not form a basis of an improved theoretical approach. The extrapolated part of the Casimir scaling function becomes surprisingly similar to that obtained experimentally \cite{chan} by Ganshin et al. Our extrapolated part can be regarded as a proposed fitting function, that appears (in the Figure 2) to agree better with the experiments \cite{chan} in D=3, than the exact mean field result of Zandi et al \cite{kardar}.

It is to be mentioned that, the mean field result of Zandi et al was improved by themselves \cite{kardar} only at $y=0$ by introducing the contribution of the critical fluctuations. The mean field result was also improved in the Ref.\cite{kardar2} by the confinement of the Goldstone modes and that of surface fluctuations.

\subsection{6. Acknowledgments}

Several useful discussions with J.K. Bhattacharjee of SNBNCBS and Kush Saha of IACS are gratefully acknowledged.

\end{document}